# A COMPLETE GRAPHICAL SOLUTION FOR UNDRAINED CYLINDRICAL CAVITY EXPANSION IN $K_0$-CONSOLIDATED MOHR-COULOMB SOIL

X. WANG, S. L. CHEN, Y. H. HAN, AND Y. ABOUSLEIMAN

**X. Wang**, Department of Civil & Environmental Engineering, Louisiana State University, Baton Rouge, LA, USA. Email: xwan134@lsu.edu

**S. L. Chen**, Department of Civil & Environmental Engineering, Louisiana State University, Baton Rouge, LA, USA (corresponding author). Email: shenglichen@lsu.edu

**Y. H. Han**, Aramco Research Center at Houston, Aramco Services Company, Houston, TX, USA. E-mail: Yanhui.Han@aramcoamericas.com

**Y. Abousleiman**, *Integrated* PoroMechanics Institute, ConocoPhillips School of Geology and Geophysics, Mewbourne School of Petroleum & Geological Engineering, The University of Oklahoma, Norman, OK, USA. Email: yabousle@ou.edu

**Abstract:** This paper develops a general and complete solution for the undrained cylindrical cavity expansion problem in non-associated Mohr-Coulomb soil under non-hydrostatic initial stress field (i.e., arbitrary $K_0$ values of the earth pressure coefficient), by expanding a unique and efficient graphical solution procedure recently proposed by Chen & Wang in 2022 for the special in situ stress case with $K_0 = 1$. The new generalized, graph-based theoretical framework contains two essential components: the geometrical analysis to track the stress path trajectory/evolution in different sectors of the deviatoric plane; and a full Lagrangian formulation of both the constitutive relationship and radial equilibrium equation to analytically determine the representative soil particle responses at the cavity surface. It is interesting to find that the cavity expansion deviatoric stress path is always composed of a series of piecewise straight lines, for all different case scenarios of $K_0$ being involved. When the cavity is sufficiently expanded, the stress path will eventually end, exclusively, in a major sextant with Lode angle $\theta$ in between $\frac{5\pi}{3}$ and $\frac{11\pi}{6}$ or on the specific line of $\theta = \frac{11\pi}{6}$, depending merely on the relative magnitude of Poisson's ratio and the friction angle. The salient advantage/feature of the present general graphical approach lies in that it can deduce the cavity expansion responses in full closed form, nevertheless being free of the limitation of the intermediacy assumption for the vertical stress and of the difficulty existing in the traditional zoning method that involves cumbersome, sequential determination of distinct Mohr-Coulomb plastic regions. Some typical results for the desired cavity expansion curves and the limit cavity pressure are presented, to investigate the impacts of soil plasticity parameters and the earth pressure coefficient on the cavity responses. The analytical closed-form solutions developed herein can be regarded as a definitive one for the undrained cavity expansion problem in classical Mohr-Coulomb materials without the approximations and simplifications in previous solutions, and will be of great value for the interpretation of pressuremeter tests in cohesive-frictional soils.

**Introduction**

Cavity expansion in elastoplastic geomaterials is one of the few well-defined initial boundary value problems involving nonlinearities that can be analytically solved in theoretical geomechanics (Selvadurai, 2007), and is a subject particularly relevant to geotechnical engineering owing to its common applications in the interpretation of pressuremeter/cone penetration test results (Yu, 2000). The pioneering work on this research topic is attributed to Gibson & Anderson (1961), who had been the first to derive both undrained and drained analytical solutions for a cylindrical cavity expanding in Mohr-Coulomb frictional soils, and to develop corresponding interpretation methods for the back calculation of soil properties from pressuremeter testing. Since then, the literature devoted to cavity expansion problems associated with different types of soil constitutive models, incorporating a diversity of material properties which range from perfect or work hardening plastic behavior to strain softening, is quite extensive. Restricting to the cohesive-frictional soils, the following are the key contributions in the development of cavity expansion analytical solutions: Ladanyi (1963); Vesic (1972); Prevost & Hoeg (1975); Hughes et al. (1977); Carter et al. (1986); Yu & Houlsby (1991); Collins et al. (1992); Papanastasiou & Durban (1997); Mantaras & Schnaid (2002); Yu & Carter (2002); Chen & Abousleiman (2018); and Carter & Yu (2022).

The above-mentioned cavity expansion analyses, based primarily on the Mohr-Coulomb model, can be broadly classified into two categories in light of the analytical solution procedures employed, i.e., the so-called similarity technique originally proposed by Hill (1950) [see Collins et al. (1992); Yu & Carter (2002); Carter & Yu (2022)] and the alternative total strain method first given by Chadwick (1959) [e.g., Yu & Houlsby (1991); Papanastasiou & Durban (1997); Chen & Abousleiman (2018)]. The common drawback of these solutions, nevertheless, lies in that the axial/vertical stress pertaining to the cylindrical case has been almost exclusively assumed to remain intermediate during the entire cavity expansion process, which unfortunately does not hold true in many instances (Florence & Schwer, 1978; Yu & Houlsby, 1991). The chief reason of introducing such an assumption in the literature was to simply avoid the corner singularity of Mohr-Coulomb type models so that closed form cavity expansion solutions can be made possible. To overcome this long-standing issue, recently a brand new graphical theoretical framework has been proposed by the authors (Chen & Abousleiman, 2022; Chen & Wang, 2022) based on a full Lagrangian formulation, which is found to be particularly powerful and efficient for rigorously solving the undrained cavity expansion problem involving the cornered Mohr-Coulomb model (Chen & Wang, 2022). The distinctive feature of the novel graphical method proposed is that, through unique and rigorous geometrical analysis, the stress path pertinent to the cavity expansion problems becomes completely trackable. This thus allows for the accurate determination of the flow rule/stiffness matrix, and eventually the development of a complete cavity expansion solution in Mohr-Coulomb soil, with the removal of the strict yet undesired intermediacy assumption for the vertical stress (Chen & Wang, 2022).

The graphical formulations in Chen & Wang (2022) are limited only to the relatively simple case of isotropic in situ stress state with the earth pressure coefficient $K_0 = 1$, which cannot apply to the general situation encountered in geotechnical practice where the out-of-plane (vertical) in situ stress may differ from the in-plane (horizontal) one (Mair & Muir Wood, 1987; Chen & Abousleiman, 2012). This paper therefore aims to generalize such a rigorous graphical approach for Mohr-Coulomb cavity expansion to achieve a full coverage of the $K_0$ value. On account of the undrained expansion conditions, it is interesting to note that the effective stress path followed by a soil particle is always composed of a series of piecewise straight lines in the deviatoric stress plane, regardless of $K_0$ greater or less than unity. This is indeed a favourable phenomenon as it essentially renders possible the derivation of the cavity expansion responses in completely closed form. For any arbitrary values of $K_0$ under consideration, the deviatoric stress paths tend to move towards the major sextant of $\frac{3\pi}{2} < \theta < \frac{11\pi}{6}$ or the specific line of $\theta = \frac{11\pi}{6}$ ($\theta$ is the Lode angle) during the continuous process of cavity expansion. And provided that the cavity is sufficiently expanded, the stress path will eventually terminate within/on the very sextant/line, depending again on the relative magnitude of Poisson's ratio and the friction angle as analogous to the case of $K_0 = 1$ (Chen & Wang, 2022). Some representative results are presented to investigate the impacts of the soil friction angle as well as the earth pressure coefficient on the essential cavity expansion curves and the limit cavity pressure. The analytical closed-form solutions developed in this work can be regarded as a definitive one for the undrained cavity expansion problem in the classical Mohr-Coulomb materials without the approximations and simplifications in previous solutions, and will in particular valuable and beneficial in geotechnical practice pertaining to the interpretation of pressuremeter tests in cohesive-frictional soils.

**General graphical method development for cylindrical cavity expansion in Mohr-Coulomb soil**

The undrained expansion problem of a cylindrical cavity in non-associated Mohr-Coulomb soil with dilation, under hydrostatic in-situ stress conditions, has recently been completely solved through the proposition of a rigorous and novel graphical solution procedure (Chen & Wang, 2022). To make the graphical theoretical framework more applicable in geotechnical practice where the vertical in situ stress is usually different from the horizontal one, it is quite natural to consider generalizing such a graph-based analytical method to cover any arbitrary values of the coefficient of earth pressure at rest, $K_0$ (defined as the ratio between the horizontal and vertical in situ effective stresses). This is exactly the objective of the present work.

The (infinitely long) cylindrical cavity expansion in elastoplastic soils is a well-defined one-dimensional boundary value problem in theoretical geomechanics. The cavity has an initial radius of $a_0$, and is expanded in an infinite saturated soil mass subjected to in-situ horizontal stress $\sigma_h$ and vertical stress $\sigma_v$, and initial pore pressure $u_0$, respectively. Let $a$ be the current radius of the cavity that corresponds to an increased internal cavity pressure $\sigma_a$. The key task in the cavity analysis is hence the determination of the cavity pressure-expansion curve between $\sigma_a$ and $a$, and of the ultimate cavity pressure $\sigma_u$ (when $a$ approaches infinity).

As already mentioned in Chen & Abousleiman (2022) and Chen & Wang (2022), the application of the graphical analysis approach for cavity expansion problem requires a full Lagrangian formulation for both the constitutive relationship and radial equilibrium equation, and needs to track the stress responses of a representative soil particle at the surface of the cavity only. Following Chen & Wang (2022), let the current soil stress state be denoted by point $\mathbf{X}(s_r, s_\theta, s_z)$ in the deviatoric, or $\pi$-plane, see Figs. 1-3 for different $K_0$ case scenarios. Here $s_r$, $s_\theta$, and $s_z$ represent the radial, tangential, and vertical deviatoric stresses, respectively, and, with the measure of the Lode angle $\theta$ as indicated in these figures, can be expressed as follows (Chen & Abousleiman, 2022)

$$s_r = \sigma'_r - p' = \frac{2}{3} q \sin(\theta + \frac{2\pi}{3}) \tag{1a}$$

$$s_\theta = \sigma'_\theta - p' = \frac{2}{3} q \sin\theta \tag{1b}$$

$$s_z = \sigma'_z - p' = \frac{2}{3} q \sin(\theta - \frac{2\pi}{3}) \tag{1c}$$

where $\sigma'_r$, $\sigma'_\theta$, and $\sigma'_z$ are, respectively, the effective principal stresses in the radial, tangential, and vertical directions (compression positive); while $p'$ and $q$, the two stress invariants of mean effective stress and deviatoric stress, are defined as

$$p' = \frac{1}{3}(\sigma'_r + \sigma'_\theta + \sigma'_z) \tag{2}$$

$$q = \sqrt{\frac{1}{2}[(\sigma'_r - \sigma'_\theta)^2 + (\sigma'_r - \sigma'_z)^2 + (\sigma'_\theta - \sigma'_z)^2]} \tag{3}$$

In addition to the current stress point $\mathbf{X}$, also specifically displayed in Figs. 1-3 are $\mathbf{X_0}(s_{r0}, s_{\theta 0}, s_{z0})$, $\mathbf{X_{ep}}(s_{r,ep}, s_{\theta,ep}, s_{z,ep})$, and $\mathbf{X_u}(s_{ru}, s_{\theta u}, s_{zu})$, which denote the projections on the deviatoric plane of the initial stress state, elastic-plastic transition stress state, and ultimate stress state, respectively. It is well known that the deviatoric stress path during the purely elastic expansion phase, i.e., segment $\mathbf{X_0 X_{ep}}$ in the figures, must be a horizontal straight line (Chen & Abousleiman, 2022; Chen & Wang, 2022). Depending on the location of the in situ stress state $\mathbf{X_0}$ (via $K_0$ obviously) relative to the two reference points $R$ and $S$ as shown in Figs. 1-3, the transition stress point $\mathrm{X_{ep}}$ may fall on different segments of $AB$, $VA$, and $BT$ that pertain to three plane faces of the yield surface situated in sextants $\frac{3\pi}{2} \leq \theta \leq \frac{11\pi}{6}$, $\frac{7\pi}{6} \leq \theta \leq \frac{3\pi}{2}$, and $-\frac{\pi}{6}$ (or $\frac{11\pi}{6}$) $\leq \theta \leq \frac{\pi}{6}$, respectively. Note that $R$ and $S$ are such chosen that $RA$ and $SB$ parallel the $x$ axis in the deviatoric plane, while $V$ and $T$ are the major and minor vertice. Three distinct case scenarios corresponding to $K_{0(R)} < K_0 < K_{0(S)}$, $K_{0(V)} < K_0 < K_{0(R)}$, and $K_{0(S)} < K_0 < K_{0(T)}$, therefore, will be analyzed separately in the development of the generalized graphical solution procedure. Here $K_{0(R)}$, $K_{0(S)}$, $K_{0(V)}$, and $K_{0(T)}$ denote the $K_0$ values corresponding to the specific points $R$, $S$, $V$, and $T$, which are given by

$$K_{0(R)} = \frac{1-(c/\sigma'_v)\cos\phi}{1+\sin\phi} \tag{4a}$$

$$K_{0(S)} = \frac{1+(c/\sigma'_v)\cos\phi}{1-\sin\phi} \tag{4b}$$

$$K_{0(V)} = \frac{1-\sin\phi-2(c/\sigma'_v)\cos\phi}{1+\sin\phi} \tag{4c}$$

$$K_{0(T)} = \frac{1+\sin\phi+2(c/\sigma_v')\cos\phi}{1-\sin\phi} \tag{4d}$$

where $c$ is the cohesion; $\phi$ is the soil frictional angle; and $\sigma_v' = \sigma_v - u_0$ denotes the initial vertical effective stress.

### (A) CASE I: $K_{0(R)} < K_0 < K_{0(S)}$

As can be clearly seen from the geometry of Fig. 1, in this first case the end point of the horizontal elastic stress path, $\mathbf{X_{ep}}$, resides on segment $AB$ and thus in the sextant of $\frac{3\pi}{2} < \theta < \frac{11\pi}{6}$ where the effective radial, tangential, and vertical principal stresses satisfy $\sigma_r' > \sigma_z' > \sigma_\theta'$. The yield function $F_{r\theta}$ and potential function $P_{r\theta}$ pertaining to point $\mathbf{X_{ep}}$, both only dependent of the major stress $\sigma_r'$ and minor stress $\sigma_\theta'$, can be expressed as

$$F_{r\theta}(\sigma_r', \sigma_\theta') = \frac{\sigma_r' - \sigma_\theta'}{2} - \frac{\sigma_r' + \sigma_\theta'}{2}\sin\phi - c\cos\phi \tag{5}$$

$$P_{r\theta}(\sigma_r', \sigma_\theta') = \frac{\sigma_r' - \sigma_\theta'}{2} - \frac{\sigma_r' + \sigma_\theta'}{2}\sin\psi + \text{const.} \tag{6}$$

where $\psi$ is the soil dilation angle.

Apparently the above two equations defining the yield and potential surfaces are the same as those involved in the specific case of $K_0 = 1$, which corresponds to hydrostatic in situ stress conditions and has been thoroughly investigated in Chen & Wang (2022). Recall that in the formulations of Chen & Wang (2022), the orientation of the plastic stress path in the deviatoric plane (or equivalently, the effective stress increment ratio of $D\sigma_r': D\sigma_\theta': D\sigma_z'$) is found to be only controlled by the values of Poisson's ratio $\nu$ and fiction angle $\phi$ and to remain a simple straight line, whenever the current stress point $\mathbf{X}$ in still located inside the sector of $\sigma_r' > \sigma_z' > \sigma_\theta'$. Due to the fact that the elastic-plastic transition stress point $\mathbf{X_{ep}}$ in the present case of $K_{0(R)} < K_0 < K_{0(S)}$ is already and always situated in this desired sector, it can be straightforwardly deduced that the deviatoric stress path must possess the same straight line characteristics (within the sextant $\sigma_r' >$ $\sigma_z' > \sigma_\theta'$) as with the $K_0 = 1$ case. The pertinent constitutive relationship, hence, can be obtained following Eqs. (5) and (6) as

$$\begin{Bmatrix} D\sigma_r' \\ D\sigma_\theta' \\ D\sigma_z' \end{Bmatrix} = \frac{1}{\Delta} \begin{bmatrix} b_{11} & b_{12} & b_{13} \\ b_{21} & b_{22} & b_{23} \\ b_{31} & b_{32} & b_{33} \end{bmatrix} \begin{Bmatrix} D\varepsilon_r \\ D\varepsilon_\theta \\ D\varepsilon_z \end{Bmatrix} \quad (7)$$

where $D\sigma_r'$, $D\sigma_\theta'$, $D\sigma_z'$ and $D\varepsilon_r$, $D\varepsilon_\theta$, $D\varepsilon_z$ are the three effective stress and strain increments in $r$, $\theta$, and $z$ directions, respectively; $\Delta$ and $b_{ij}$ ($i, j = 1, 2, 3$) are all constants depending explicitly on $\nu$, $\phi$, $\psi$, and shear modulus $G$, which have the same definitions as those in Chen & Wang (2022).

Now introduce the radial equilibrium equation in the form of Lagrangian description (Chen & Abousleiman, 2013, 2022)

$$\frac{D\sigma_a}{D\xi} = \frac{1-\xi}{2\xi - \xi^2} (\sigma_a' - \sigma_{\theta,a}') \quad (8)$$

where $\sigma_a'$ and $\sigma_{\theta,a}'$ denote the effective radial and tangential stresses at cavity wall, respectively; and $\xi$ is known as an auxiliary variable defined as $\xi = \frac{a - a_0}{a}$ (Chen & Abousleiman, 2013, 2022). It should be noted that Eq. (8), converted from the Eulerian-based equation via the pivotal auxiliary variable $\xi$, is related to the stress state for a specific point at the cavity surface only.

Following a similar procedure used in Chen & Wang (2022), i.e., analytically solving $\sigma_r'$, $\sigma_\theta'$, and $\sigma_z'$ from Eq. (7) for a representative soil particle at the cavity wall under undrained conditions ($D\varepsilon_r + D\varepsilon_\theta = D\varepsilon_z = 0$), and on substituting in Eq. (8), one obtains the following closed-form expressions for the radial effective stress $\sigma_a'$, cavity pressure $\sigma_a$, and the corresponding pore pressure $u_a$:

$$\sigma_a' = \frac{b_{11} - b_{12}}{\Delta} \ln\left(\frac{a}{a_{ep}}\right) + \sigma_{r,ep}' \quad (9)$$

$$\sigma_a = f_{r\theta}(a) - f_{r\theta}(a_{ep}) + \sigma_{r,ep}' + u_0 \quad (10)$$

$$u_a = \sigma_a - \sigma'_a = f_{r\theta}(a) - f_{r\theta}(a_{ep}) - \frac{b_{11}-b_{12}}{\Delta}\ln\left(\frac{a}{a_{ep}}\right) + u_0 \tag{11}$$

with

$$f_{r\theta}(a) = A_{r\theta}\left\{\frac{1}{2}\ln\left(\frac{a}{a_0}\right)\ln\left(\frac{a^2}{a_0^2}-1\right) - \left[\ln\left(\frac{a}{a_0}\right)\right]^2 - \frac{1}{4}\text{Li}_2\left(\frac{a_0^2}{a^2}\right)\right\} + \frac{A_{r\theta}\ln B + C}{2}\ln\left(1-\frac{a_0^2}{a^2}\right) \tag{12}$$

where $\text{Li}_2$ represents the polylogarithm function of order 2; $A_{r\theta} = \frac{(b_{11}-b_{12}-b_{21}+b_{22})}{\Delta}$; $B = 1 - \frac{\sigma'_{r,ep}-\sigma'_h}{2G}$; $C = \sigma'_{r,ep} - \sigma'_{\theta,ep}$; $a_{ep} = \frac{a_0}{B}$ is the increased cavity radius at the transaction from elastic to plastic straining; while $\sigma'_h = \sigma_h - u_0$ denotes the initial horizontal effective stress, and $\sigma'_{r,ep}$ and $\sigma'_{\theta,ep}$ pertain to the effective radial and tangential stresses at the transition point $\mathbf{X_{ep}}$ which, along with the third vertical stress component $\sigma'_{z,ep}$, can be readily determined as

$$\sigma'_{r,ep} = (1+\sin\phi)K_0\sigma'_v + c\cos\phi, \quad \sigma'_{\theta,ep} = (1-\sin\phi)K_0\sigma'_v - c\cos\phi, \quad \sigma'_{z,ep} = \sigma'_v \tag{13}$$

It is favorable to see that Eqs. (9)-(12) are symbolically almost identical to those corresponding to the case of $K_0 = 1$ (see Chen & Wang, 2022), which is expected and in turn indicates that the graphical formulation/solution previously developed for the cavity expansion in Mohr-Coulomb soil under isotropic in situ stress state is indeed equally applicable for the more general values of $K_0$ ranging from $K_{0(R)}$ to $K_{0(S)}$. One therefore can further conceive that if the condition of $2\nu \geq 1 - \sin\phi$ holds true, then the plastic stress path $\mathbf{X_{ep}X}$ will be strictly confined within the major sextant of $\sigma'_r > \sigma'_z > \sigma'_\theta$ during the whole cavity expansion process, until the ultimate state $\mathbf{X_u}$ is arrived at (see the shaded area as illustrated in Fig. 1). Under such circumstance, the limiting cavity pressure $\sigma_u$ can be obtained from the above Eq. (10) as

$$\sigma_u = \lim_{a\to\infty}\sigma_a = \frac{A_{r\theta}}{4}\text{Li}_2(B^2) - \frac{C}{2}\ln(1-B^2) + \sigma'_{r,ep} + u_0 \tag{14}$$

The above description applies to the case of $2\nu \geq 1 - \sin\phi$ for the entire stress path from point $\mathbf{X_{ep}}$, through $\mathbf{X}$ to the ultimate state $\mathbf{X_u}$, and to the alternative case $2\nu < 1 - \sin\phi$ as well (except for Eq. (14)) as long as the stress state **X** is still located in the sextant $\sigma'_r > \sigma'_z > \sigma'_\theta$. However, when $2\nu < 1 - \sin\phi$, the stress path $\mathbf{X_{ep}X}$ will be oriented in such a way that it eventually hits the line $\theta = \frac{11\pi}{6}$ at a certain point $\mathbf{X_{pr}}$ and then stick to that line afterwards until the ultimate point $\mathbf{X_u}$ is reached, see the trajectory $\mathbf{X_{ep}XX_{pr}X'X_u}$ as shown in Fig. 1. The fact that the segment $\mathbf{X_{pr}X'X_u}$ shall be in alignment with the projected $s_r$ axis can again be proved following exactly the same graphical analysis provided in Chen & Wang (2022). And for this corner loading condition with the current stress state $\mathbf{X'}$ lying on the common edge of the previously defined yield/potential surfaces $F_{r\theta}/P_{r\theta}$ and the following ones of $F_{rz}/P_{rz}$ (in the sextant $\sigma'_r > \sigma'_\theta > \sigma'_z$):

$$F_{rz}(\sigma'_r, \sigma'_z) = \frac{\sigma'_r - \sigma'_z}{2} - \frac{\sigma'_r + \sigma'_z}{2}\sin\phi - c\cos\phi \tag{15}$$

$$P_{rz}(\sigma'_r, \sigma'_z) = \frac{\sigma'_r - \sigma'_z}{2} - \frac{\sigma'_r + \sigma'_z}{2}\sin\psi + \text{const.} \tag{16}$$

the corresponding stress and pore pressure analysis/responses for a soil particle at the cavity wall that were presented in Chen & Wang (2022) must also still be valid. Hence, one has similarly

$$\sigma'_a = \frac{2(1+m)n}{\Delta_r}\ln\left(\frac{a}{a_{pr}}\right) + \sigma'_{r,pr} \tag{17}$$

$$\sigma_a = g(a) - g(a_{pr}) + \sigma'_{r,pr} + u_{pr} \tag{18}$$

$$u_a = \sigma_a - \sigma'_a = g(a) - g(a_{pr}) - \frac{2(1+m)n}{\Delta_r}\ln\left(\frac{a}{a_{pr}}\right) + u_{pr} \tag{19}$$

$$\sigma_u = \lim_{a\to\infty}\sigma_a = \frac{mn}{\Delta_r}\text{Li}_2\left(\frac{B^2}{R_{pr}^2}\right) - \frac{A_{r\theta}\ln R_{pr}+C}{2}\ln\left(1-\frac{B^2}{R_{pr}^2}\right) + \sigma'_{r,pr} + u_{pr} \tag{20}$$

with

$$g(a) = \frac{4mn}{\Delta_r}\left\{\frac{1}{2}\ln\left(\frac{a}{a_0}\right)\ln\left(\frac{a^2}{a_0^2}-1\right) - \left[\ln\left(\frac{a}{a_0}\right)\right]^2 - \frac{1}{4}\text{Li}_2\left(\frac{a_0^2}{a^2}\right)\right\}$$

$$+ \frac{1}{2}\left[A_{r\theta}\ln R_{pr} + \frac{4mn}{\Delta_r}\ln\left(\frac{B}{R_{pr}}\right) + C\right]\ln\left(1-\frac{a_0^2}{a^2}\right) \tag{21}$$

where $m = \sin\phi$; $n = \sin\psi$; $\Delta_r = \frac{1}{2G(1+v)}[3 - 6v + (2v - 1)(m + n) + (3 + 2v)mn]$; $R_{pr} = \exp\left\{\frac{\Delta(\sigma'_{z,ep} - \sigma'_{\theta,ep})}{b_{21} - b_{22} - b_{31} + b_{32}}\right\}$; $a_{pr} = a_{ep}R_{pr}$ represents the cavity radius corresponding to the intersection stress point $\mathbf{X_{pr}}$; while $\sigma'_{r,pr}$ and $u_{pr}$ are the associated effective radial stress and pore pressure, which can be readily determined through substitution of $a_{pr}$ for $a$ in Eqs. (9) and (11), respectively.

### (B) CASE II: $\mathbf{K_{0(V)} < K_0 < K_{0(R)}}$

For the case of $K_{0(V)} < K_0 < K_{0(R)}$, the initial stress state $\mathbf{X_0}$ lies on the segment $RV$ of the triaxial compression line with $\theta = \frac{7\pi}{6}$, refer to Fig. 2. The elastic-plastic transition point $\mathbf{X_{ep}}$, as a result of the horizontal direction of the pure elastic stress path in the deviatoric plane, now shifts upwards to the sextant $\frac{7\pi}{6} < \theta < \frac{3\pi}{2}$ characterized by $\sigma'_z > \sigma'_r > \sigma'_\theta$. Accordingly, the two planar Mohr-Coulomb yield and potential surfaces passing through $\mathbf{X_{ep}}$ are related only to the major stress $\sigma'_z$ and minor stress $\sigma'_\theta$, which are governed by

$$F_{z\theta}(\sigma'_z, \sigma'_\theta) = \frac{\sigma'_z - \sigma'_\theta}{2} - \frac{\sigma'_z + \sigma'_\theta}{2}\sin\phi - c\cos\phi \tag{22}$$

$$P_{z\theta}(\sigma'_z, \sigma'_\theta) = \frac{\sigma'_z - \sigma'_\theta}{2} - \frac{\sigma'_z + \sigma'_\theta}{2}\sin\psi + \text{const.} \tag{23}$$

Combining the yield criterion (22) with the elastic cavity expansion solution (Yu, 2000; Chen & Abousleiman, 2012), the effective stress components at point $\mathbf{X_{ep}}$ (on segment $VA$ for the present case) can be easily shown to be

$$\sigma'_{r,ep} = (2K_0 - \frac{1-\sin\phi}{1+\sin\phi})\sigma'_v + \frac{2c\cos\phi}{1+\sin\phi}, \quad \sigma'_{\theta,ep} = \frac{1-\sin\phi}{1+\sin\phi}\sigma'_v - \frac{2c\cos\phi}{1+\sin\phi}, \quad \sigma'_{z,ep} = \sigma'_v \tag{24}$$

Beyond point $\mathbf{X_{ep}}$, it is found that the plastic deviatoric stress path remains a straight line within the sector of $\frac{7\pi}{6} < \theta < \frac{3\pi}{2}$ (see segment $\mathbf{X_{ep}}\hat{\mathbf{X}}$ in Fig. 2) and tends to approach the negative $s_\theta$ axis with the Lode angle $\theta = \frac{3\pi}{2}$. The latter of the downward trend of the movement can be straightforwardly illustrated in a graphical manner as follows. Note that for the undrained cavity expansion problem, the deviator of incremental strain vector, $D\boldsymbol{e}$, is always directed in the horizontal sense in the deviatoric strain plane (Chen & Abousleiman, 2022; Chen & Wang, 2022). While $D\boldsymbol{e}$, on the other hand, can be decomposed into two components in the $\pi$-plane: the elastic deviatoric component $D\boldsymbol{e}^e$ along the direction of incremental stress (i.e., $D\boldsymbol{e}^e \parallel \mathbf{X}_{ep}\widehat{\mathbf{X}}$), and the plastic component $D\boldsymbol{e}^p$ normal to the devitoric cross-section of the potential surface $P_{z\theta}$ (i.e., the dashed line passing through $\widehat{\mathbf{X}}$). Since $D\boldsymbol{e}^p$ is already pointing upward and to the right, from the geometry of Fig. 2, it then becomes obvious that $D\boldsymbol{e}^e$ must orient in a direction below the horizontal line to generate a resultant deviatoric strain increment $D\boldsymbol{e} \parallel \boldsymbol{O}x$. This proves that $\mathbf{X}_{ep}\widehat{\mathbf{X}}$ in parallel to $D\boldsymbol{e}^e$ has to move towards the line $\theta = \frac{3\pi}{2}$, with the continued expansion of the cavity.

To verify the straight line nature of the stress path segment $\mathbf{X}_{ep}\widehat{\mathbf{X}}$ developed in the sextant $\frac{7\pi}{6} < \theta < \frac{3\pi}{2}$, one may resort to the elastoplastic constitutive relationship for a representative stress state $\widehat{\mathbf{X}}$ lying on this very sector. By making use of Eqs. (22) and (23), this relationship can be written as (Chen & Han, 1988; Yu, 1994)

$$\begin{Bmatrix} D\sigma'_r \\ D\sigma'_\theta \\ D\sigma'_z \end{Bmatrix} = \frac{1}{\Delta}\begin{bmatrix} b_{33} & b_{32} & b_{31} \\ b_{23} & b_{22} & b_{21} \\ b_{13} & b_{12} & b_{11} \end{bmatrix}\begin{Bmatrix} D\varepsilon_r \\ D\varepsilon_\theta \\ D\varepsilon_z \end{Bmatrix} \tag{25}$$

where $\Delta$ and $b_{ij}$ have been defined previously.

With the aid of $D\varepsilon_r = -D\varepsilon_\theta$ and $D\varepsilon_z = 0$ for the undrained plane-strain expansion conditions, Eq. (25) leads to the following effective stress increment ratio

$$\begin{aligned} D\sigma'_r : D\sigma'_\theta : D\sigma'_z &= (b_{33} - b_{32}):(b_{23} - b_{22}):(b_{13} - b_{12}) \\ &= 2[1 - 2\nu + (\nu + \sin\phi + \nu\sin\phi)\sin\psi] \\ &\quad :(1 - \sin\phi)(2\nu + \sin\psi - 1):(1 + \sin\phi)(2\nu + \sin\psi - 1) \end{aligned} \tag{26}$$

The above equation clearly indicates that the orientation of the effective stress increment vector $D\boldsymbol{\sigma}' = \{D\sigma_r', D\sigma_\theta', D\sigma_z'\}^T$ at point $\hat{\mathbf{X}}$ is only dependent on the three constant parameters $\nu$, $\phi$, and $\psi$, but not on the current stress state of the soil itself. Therefore, the stress path in the sextant $\frac{7\pi}{6} < \theta < \frac{3\pi}{2}$ indeed turns out to be a straight line with the direction defined by Eq. (26). Nevertheless, it should be noted that here the orientation of $D\boldsymbol{\sigma}'$ is also affected by the dilation angle $\psi$ in addition to the Poisson's ratio $\nu$ and friction angle $\phi$; the latter are the two only contributing soil parameters that controls the stress path when it is located in the major sextant $\frac{3\pi}{2} < \theta < \frac{11\pi}{6}$, as described in the preceding subsection.

Now substituting the large strain increments definition of $D\varepsilon_r = -D\varepsilon_\theta = \frac{Dr}{r}$ (where $Dr$ denotes the infinitesimal change in the radial position) into Eq. (25), and replacing $r$ by $a$ followed by integration to track the stress responses for a soil particle at the cavity wall, the three effective stress components $\sigma_a'$, $\sigma_{\theta,a}'$, and $\sigma_{z,a}'$ in the radial, tangential, and vertical directions can be analytically obtained as

$$\sigma_a' = \frac{b_{33}-b_{32}}{\Delta} \ln \frac{a}{a_{ep}} + \sigma_{r,ep}' \tag{27a}$$

$$\sigma_{\theta,a}' = \frac{b_{23}-b_{22}}{\Delta} \ln \frac{a}{a_{ep}} + \sigma_{\theta,ep}' \tag{27b}$$

$$\sigma_{z,a}' = \frac{b_{13}-b_{12}}{\Delta} \ln \frac{a}{a_{ep}} + \sigma_{z,ep}' \tag{27c}$$

Here $\sigma_{r,ep}'$, $\sigma_{\theta,ep}'$, and $\sigma_{z,ep}'$ are already given in Eq. (24), which should not be confused with those presented in Eq. (13) for the previous case of $K_{0(R)} < K_0 < K_{0(S)}$; and $a_{ep}$, the cavity radius corresponding to the initial yielding, can still be found from the expression $a_{ep} = \frac{a_0}{B}$, but with the involved $\sigma_{r,ep}'$ again being calculated by means of Eq. (24).

Having known the effective stress components, the solution procedure for determining the cavity pressure $\sigma_a$ and the related pore pressure $u_a$ is completely analogous to the previous one for Case I. The equilibrium condition of Eq. (8) remains unchanged, which, with the substitution of Eqs. (27a) and (27b), leads to

$$\frac{D\sigma_a}{Da} = \frac{1}{a[(a/a_0)^2 - 1]} \left\{ A_{z\theta} \ln\left[B\left(\frac{a}{a_0}\right)\right] + C \right\} \tag{28}$$

where $A_{z\theta} = \frac{(b_{33} - b_{32} - b_{23} + b_{22})}{\Delta}$.

The solution of the differential equation is

$$\sigma_a = f_{z\theta}(a) - f_{z\theta}(a_{ep}) + \sigma'_{r,ep} + u_0 \tag{29}$$

where

$$f_{z\theta}(a) = A_{z\theta} \left\{ \frac{1}{2} \ln\left(\frac{a}{a_0}\right) \ln\left(\frac{a^2}{a_0^2} - 1\right) - \left[\ln\left(\frac{a}{a_0}\right)\right]^2 - \frac{1}{4} \text{Li}_2\left(\frac{a_0^2}{a^2}\right) \right\} + \frac{A_{z\theta} \ln B + C}{2} \ln\left(1 - \frac{a_0^2}{a^2}\right) \tag{30}$$

Eq. (11) becomes

$$u_a = \sigma_a - \sigma'_a = f_{z\theta}(a) - f_{z\theta}(a_{ep}) - \frac{b_{33} - b_{32}}{\Delta} \ln\left(\frac{a}{a_{ep}}\right) + u_0 \tag{31}$$

It is understood that the analytical solutions derived above, i.e., Eqs. (27a)-(27c), and (29), are valid for any stress point $\hat{\mathbf{X}}$ positioned in the sextant $\frac{7\pi}{6} < \theta < \frac{3\pi}{2}$, before the attainment of the intersection point $\mathbf{X_{p\theta}}$ between the straight stress path $\mathbf{X_{ep}X_{p\theta}}$ and the negative $s_\theta$ axis ($\theta = \frac{3\pi}{2}$). After arriving at the common edge of $\theta = \frac{3\pi}{2}$ of the two adjacent yield surfaces $F_{z\theta}$ and $F_{r\theta}$, the stress path theoretically may further develop in three different ways. It can bounce back to the upper sextant $\frac{7\pi}{6} < \theta < \frac{3\pi}{2}$, remain on the line $\theta = \frac{3\pi}{2}$, or move across into the major sextant of $\frac{3\pi}{2} < \theta < \frac{11\pi}{6}$, see the corresponding segments of $\mathbf{X_{p\theta}\overline{X}}$, $\mathbf{X_{p\theta}\widetilde{X}}$, and $\mathbf{X_{p\theta}X}$ as shown in Fig. 2. Following basically the same graphical analysis of the stress path as described in Chen & Wang (2022) for the case of $2\nu < 1 - \sin\phi$, however, one may easily deduce that the only possible stress path beyond point $\mathbf{X_{p\theta}}$ must be the third one of $\mathbf{X_{p\theta}X}$. This is because the first path of $\mathbf{X_{p\theta}\overline{X}}$ will evidently contradict the downward movement requirement (towards the $s_\theta$ axis), as has already been graphically interpreted above. While the second path $\mathbf{X_{p\theta}\widetilde{X}}$ along the negative $s_\theta$ axis will result that the elastic strain increment $De^e$, as well as the two components of the plastic strain increment $De^p$ pertaining to the potential functions $P_{z\theta}$ and $P_{r\theta}$, all orient in a direction above the horizontal line, which therefore is in contradiction with the basic requirement of $De \parallel \mathbf{O}x$ for the undrained cavity expansion deformation.

Once the stress state subsequently enters into the sector of $\frac{3\pi}{2} < \theta < \frac{11\pi}{6}$ in the deviatoric plane, the vertical effective stress $\sigma'_z$ turns out to be the intermediate principal stress, i.e., $\sigma'_r > \sigma'_z > \sigma'_\theta$. In this situation the graphical analysis method described earlier for the stress path and cavity responses for the first case of $K_{0(R)} < K_0 < K_{0(S)}$ must be equally applicable, with the difference that the initial conditions involved need to be associated with point $\mathbf{X_{p\theta}}$ instead of $\mathbf{X_{ep}}$ as in Case I. If $2\nu \geq 1 - \sin\phi$, the stress path $\mathbf{X_{p\theta}X}$ will again remain strictly in the major sector of $\frac{3\pi}{2} < \theta < \frac{11\pi}{6}$ all the way down to the ultimate state $\mathbf{X_u}$. The corresponding equations for $\sigma_a$, $u_a$, and $\sigma_u$ are reduced to

$$\sigma_a = \tilde{f}_{r\theta}(a) - \tilde{f}_{r\theta}(a_{p\theta}) + \sigma'_{r,p\theta} + u_{p\theta} \tag{32}$$

$$u_a = \tilde{f}_{r\theta}(a) - \tilde{f}_{r\theta}(a_{p\theta}) - \frac{b_{11}-b_{12}}{\Delta}\ln\left(\frac{a}{a_{p\theta}}\right) + u_{p\theta} \tag{33}$$

$$\sigma_u = \lim_{a\to\infty}\sigma_a = \frac{A_{r\theta}}{4}\text{Li}_2\left(\frac{B^2}{R_{p\theta}^2}\right) - \frac{A_{\theta z}\ln R_{p\theta}+C}{2}\ln\left(1 - \frac{B^2}{R_{p\theta}^2}\right) + \sigma'_{r,p\theta} + u_{p\theta} \tag{34}$$

with

$$\tilde{f}_{r\theta}(a) = A_{r\theta}\left\{\frac{1}{2}\ln\left(\frac{a}{a_0}\right)\ln\left(\frac{a^2}{a_0^2}-1\right) - \left[\ln\left(\frac{a}{a_0}\right)\right]^2 - \frac{1}{4}\text{Li}_2\left(\frac{a_0^2}{a^2}\right)\right\}$$

$$+ \frac{A_{r\theta}\ln(B/R_{p\theta}) + A_{\theta z}\ln R_{p\theta} + C}{2}\ln\left(1 - \frac{a_0^2}{a^2}\right) \qquad (35)$$

where $R_{p\theta} = \exp\left\{\frac{\Delta(\sigma'_{z,ep} - \sigma'_{r,ep})}{b_{12} - b_{13} - b_{32} + b_{33}}\right\}$; $a_{p\theta} = R_{p\theta}a_{ep}$ represents the cavity radius corresponding to the intersection point $\mathbf{X_{p\theta}}$; while $\sigma'_{r,p\theta}$ and $u_{p\theta}$ denote the related effective radial stress and pore pressure, which can be computed, respectively, from Eqs. (27a) and (31) by replacing $a$ with $a_{p\theta}$.

In contrast, if the condition $2\nu < 1 - \sin\phi$ is satisfied, then the above equations (32) and (33) will still be valid for the calculation of cavity responses, as long as $\sigma_a < \sigma'_{r,pr} + u_{pr}$ (or $a < a_{pr}$) where $a_{pr}$, $\sigma'_{r,pr}$, and $u_{pr}$ denote the cavity radius and the corresponding cavity pressure and pore pressure at the intersection point $\mathbf{X_{pr}}$ with the $s_r$ axis (see Fig. 2). Otherwise, the stress path shall end up with lying on the projected $s_r$ axis, following the line $\mathbf{X_{p\theta}XX_{pr}X'X_u}$ as shown in the same figure. The cavity expansion solution therefore will be an analogue to the expressions in Eqs. (18)-(21), which takes a slightly different form as follow:

$$\sigma_a = \tilde{g}(a) - \tilde{g}(a_{pr}) + \sigma'_{r,pr} + u_{pr} \qquad (36)$$

$$u_a = \tilde{g}(a) - \tilde{g}(a_{pr}) - \frac{2(1+m)n}{\Delta_r}\ln\left(\frac{a}{a_{pr}}\right) + u_{pr} \qquad (37)$$

$$\sigma_u = \lim_{a \to \infty}\sigma_a = \frac{mn}{\Delta_r}\text{Li}_2\left(\frac{B^2}{R_{pr}^2 R_{p\theta}^2}\right) - \frac{A_{r\theta}\ln R_{pr} + A_{z\theta}\ln R_{p\theta} + C}{2}\ln\left(1 - \frac{B^2}{R_{pr}^2 R_{p\theta}^2}\right) + \sigma'_{r,pr} + u_{pr} \qquad (38)$$

with

$$\tilde{g}(a) = \frac{4mn}{\Delta_r}\left\{\frac{1}{2}\ln\left(\frac{a}{a_0}\right)\ln\left(\frac{a^2}{a_0^2} - 1\right) - \left[\ln\left(\frac{a}{a_0}\right)\right]^2 - \frac{1}{4}\text{Li}_2\left(\frac{a_0^2}{a^2}\right)\right\}$$

$$+ \frac{A_{r\theta}\ln R_{pr} + A_{z\theta}\ln R_{p\theta} + (4mn/\Delta_r)\ln[B/(R_{pr}R_{p\theta})] + C}{2}\ln\left(1 - \frac{a_0^2}{a^2}\right) \qquad (39)$$

where $R_{pr} = \exp\left\{\frac{\Delta(\sigma'_{z,p\theta} - \sigma'_{\theta,p\theta})}{b_{21} - b_{22} - b_{31} + b_{32}}\right\}$, with $\sigma'_{z,p\theta}$ and $\sigma'_{\theta,p\theta}$ obtained by putting $a = a_{p\theta} = R_{p\theta}a_{ep}$ in Eqs. (27b) and (27c); $a_{pr} = R_{pr}a_{p\theta}$ represents the cavity radius pertaining to point $\mathbf{X_{pr}}$; while $\sigma'_{r,pr}$ and $u_{pr}$ denoting the related effective radial stress and pore pressure should be determined from Eqs. (32) and (33) with $a$ set to be $a_{pr}$.

**(C) CASE III: $K_{0(S)} < K_0 < K_{0(T)}$**

For this last category of $K_0$ values in between $K_{0(S)}$ and $K_{0(T)}$, the in-situ stress state $\mathbf{X_0}$ has moved downward to the portion $ST$ of the triaxial extension line with $\theta = \frac{\pi}{6}$. As a consequence, the corresponding elastic-plastic transition point $\mathbf{X_{ep}}$, as shown in Fig. 3, now falls on segment $BT$ of the deviatoric cross section of the yield surface $F_{rz}$. It is therefore expected that the developed stress path span two adjacent sextants of $-\frac{\pi}{6} < \theta < \frac{\pi}{6}$ and $\frac{3\pi}{2} < \theta < \frac{11\pi}{6}$, and will again either be confined to the latter (major) sector with $\sigma'_r > \sigma'_z > \sigma'_\theta$ or end up with lying on the $s_r$ axis with $\sigma'_\theta = \sigma'_z$ (depending on the relative magnitudes of $\nu$ and $\phi$), see the respective stress paths $\mathbf{X_0 X_{ep} \check{X} X_{pr} X X_u}$ and $\mathbf{X_0 X_{ep} \check{X} X_{pr} X' X_u}$ presented in Fig. 3.

As similar with Case II: $K_{0(V)} < K_0 < K_{0(R)}$, the solution procedure will begin with analyzing the elastoplastic cavity responses pertaining to the stress path segment $\mathbf{X_{ep} \check{X} X_{pr}}$ located in the sector of $-\frac{\pi}{6} < \theta < \frac{\pi}{6}$. For any stress point $\check{\mathbf{X}}$ falling within this sector, the constitutive relationship, by means of Eqs. (15) and (16), transforms into

$$\begin{Bmatrix} D\sigma'_r \\ D\sigma'_\theta \\ D\sigma'_z \end{Bmatrix} = \frac{1}{\Delta} \begin{bmatrix} b_{11} & b_{13} & b_{12} \\ b_{31} & b_{33} & b_{32} \\ b_{21} & b_{23} & b_{22} \end{bmatrix} \begin{Bmatrix} D\varepsilon_r \\ D\varepsilon_\theta \\ D\varepsilon_z \end{Bmatrix} \quad (40)$$

where $\Delta$ and $b_{ij}$ are only dependent of $\nu$, $\phi$, $\psi$, and $G$, already defined.

From Eq. (40), one obtains

$$D\sigma'_r : D\sigma'_\theta : D\sigma'_z = (b_{11} - b_{13}) : (b_{31} - b_{33}) : (b_{21} - b_{23})$$
$$= (1 + \sin\phi)(1 - 2\nu + \sin\psi) :$$
$$2[(\nu - \sin\phi - \nu\sin\phi)\sin\psi - 1 + 2\nu] : (1 - \sin\phi)(1 - 2\nu + \sin\psi) \quad (41)$$

which again clearly demonstrates that the stress path developed in the sextant $-\frac{\pi}{6} < \theta < \frac{\pi}{6}$ during the early elastoplastic deformation phase of the cylindrical cavity expansion must be a straight line.

The equations analogous to Eqs. (27a)-(27c), obtained by integrating Eq. (40) with respect to $a$, are

$$\sigma'_a = \frac{b_{11}-b_{13}}{\Delta} \ln \frac{a}{a_{ep}} + \sigma'_{r,ep} \tag{42a}$$

$$\sigma'_{\theta,a} = \frac{b_{31}-b_{33}}{\Delta} \ln \frac{a}{a_{ep}} + \sigma'_{\theta,ep} \tag{42b}$$

$$\sigma'_{z,a} = \frac{b_{21}-b_{23}}{\Delta} \ln \frac{a}{a_{ep}} + \sigma'_{z,ep} \tag{42c}$$

where

$$\sigma'_{r,ep} = \frac{1+\sin\phi}{1-\sin\phi}\sigma'_v + \frac{2c\cos\phi}{1-\sin\phi}, \quad \sigma'_{\theta,ep} = (2K_0 - \frac{1+\sin\phi}{1-\sin\phi})\sigma'_v - \frac{2c\cos\phi}{1-\sin\phi}, \quad \sigma'_{z,ep} = \sigma'_v \tag{43}$$

Note that for point $\check{X}$, the deviatoric plastic strain increment $De^p$ is oriented downward and to the right (see Fig. 3). It therefore can be readily proved graphically that the straight stress path $X_{ep}\check{X}X_{pr}$ must be in a direction above the horizontal line, i.e., inclined towards the $s_r$ axis, to accommodate the horizontal resultant $De$ under the undrained conditions. Furthermore, the analytical solutions for the cavity pressure and pore pressure pertaining to this segment of stress path may be found following the procedure previously outlined for point $\hat{X}$ in the sextant of $\frac{7\pi}{6} < \theta < \frac{3\pi}{2}$, giving

$$\sigma_a = f_{rz}(a) - f_{rz}(a_{ep}) + \sigma'_{r,ep} + u_0 \tag{44}$$

$$u_a = \sigma_a - \sigma'_a = f_{rz}(a) - f_{rz}(a_{ep}) - \frac{b_{11}-b_{13}}{\Delta} \ln\left(\frac{a}{a_{ep}}\right) + u_0 \tag{45}$$

where

$$f_{rz}(a) = A_{rz}\left\{\frac{1}{2}\ln\left(\frac{a}{a_0}\right)\ln\left(\frac{a^2}{a_0^2}-1\right) - \left[\ln\left(\frac{a}{a_0}\right)\right]^2 - \frac{1}{4}\text{Li}_2\left(\frac{a_0^2}{a^2}\right)\right\} + \frac{A_{rz}\ln B + C}{2}\ln\left(1-\frac{a_0^2}{a^2}\right) \tag{46}$$

with $A_{rz} = \frac{(b_{11}-b_{13}-b_{31}+b_{33})}{\Delta}$; and $a_{ep}$ has the same definition as before with the involved $\sigma'_{r,ep}$ nevertheless being determined from Eq. (43).

As $\sigma_a$ increases further, the stress path beyond point $\mathbf{X_{pr}}$ will again be controlled by the relative magnitudes of $2\nu$ and $1 - \sin\phi$. If $2\nu \geq 1 - \sin\phi$, the stress path tends to cross the $s_r$ axis and then follows a typical straight line $\mathbf{X_{pr}XX_u}$ that will be fully contained in the major sextant of $\frac{3\pi}{2} < \theta < \frac{11\pi}{6}$. The analogous set of solutions for the cavity pressure/pore pressure responses can be obtained from the previous analysis by slightly changing the expressions in Eqs. (32)-(35), such that

$$\sigma_a = \hat{f}_{r\theta}(a) - \hat{f}_{r\theta}(a_{p\theta}) + \sigma'_{r,pr} + u_{pr} \tag{47}$$

$$u_a = \hat{f}_{r\theta}(a) - \hat{f}_{r\theta}(a_{pr}) - \frac{b_{11}-b_{12}}{\Delta}\ln\left(\frac{a}{a_{pr}}\right) + u_{pr} \tag{48}$$

$$\sigma_u = \lim_{a \to \infty} \sigma_a = \frac{A_{r\theta}}{4}\text{Li}_2\left(\frac{B^2}{R_{pr}^2}\right) - \frac{A_{rz}\ln R_{pr}+C}{2}\ln\left(1 - \frac{B^2}{R_{pr}^2}\right) + \sigma'_{r,pr} + u_{pr} \tag{49}$$

with

$$\hat{f}_{r\theta}(a) = A_{r\theta}\left\{\frac{1}{2}\ln\left(\frac{a}{a_0}\right)\ln\left(\frac{a^2}{a_0^2} - 1\right) - \left[\ln\left(\frac{a}{a_0}\right)\right]^2 - \frac{1}{4}\text{Li}_2\left(\frac{a_0^2}{a^2}\right)\right\}$$

$$+ \frac{A_{r\theta}\ln B/R_{pr} + A_{rz}\ln R_{pr}+C}{2}\ln\left(1 - \frac{a_0^2}{a^2}\right) \tag{50}$$

where $a_{pr} = R_{pr}a_{ep}$ and $R_{pr} = \exp\left\{\frac{\Delta(\sigma'_{z,ep}-\sigma'_{\theta,ep})}{b_{31}-b_{33}-b_{21}+b_{23}}\right\}$; while $\sigma'_{r,pr}$ and $u_{pr}$ at $a = a_{pr}$ this time should be calculated according to Eqs. (44) and (45) above.

Similarly, if $2\nu < 1 - \sin\phi$, the stress path $(\mathbf{X_{pr}X'X_u})$ will continue staying along the $s_r$ axis after the intersection point $\mathbf{X_{pr}}$ has been reached. In this case it is conceivable that the cavity/pore pressure-expansion responses be analogous to Eqs. (36) and (39), and found to be

$$\sigma_a = \hat{g}(a) - \hat{g}(a_{pr}) + \sigma'_{r,pr} + u_{pr} \tag{51}$$

$$u_a = \hat{g}(a) - \hat{g}(a_{pr}) - \frac{2(1+m)n}{\Delta_r} \ln\left(\frac{a}{a_{pr}}\right) + u_{pr} \tag{52}$$

$$\sigma_u = \lim_{a \to \infty} \sigma_a = \frac{mn}{\Delta_r} \text{Li}_2\left(\frac{B^2}{R_{pr}^2}\right) - \frac{A_{rz}\ln R_{pr}+C}{2} \ln\left(1 - \frac{B^2}{R_{pr}^2}\right) + \sigma'_{r,pr} + u_{pr} \tag{53}$$

with

$$\hat{g}(a) = \frac{4mn}{\Delta_r}\left\{\frac{1}{2}\ln\left(\frac{a}{a_0}\right)\ln\left(\frac{a^2}{a_0^2} - 1\right) - \left[\ln\left(\frac{a}{a_0}\right)\right]^2 - \frac{1}{4}\text{Li}_2\left(\frac{a_0^2}{a^2}\right)\right\}$$

$$+ \frac{A_{rz}\ln R_{pr} + (4mn/\Delta_r)\ln[B/R_{pr}] + C}{2} \ln\left(1 - \frac{a_0^2}{a^2}\right) \tag{54}$$

**Numerical results**

In this section some sample results are presented for the cavity expansion curve and the limit cavity pressure, to illustrate particularly how they are impacted by the key parameters of the soil friction angle $\phi$ and earth pressure coefficient $K_0$. Note that these curves are calculated from the explicit expressions, Eqs. (10), (14), (18), (20); Eqs. (29), (32), (34), (36), (38); and Eqs. (44), (47), (49), (51), (53) for the three cases of $K_{0(R)} < K_0 < K_{0(S)}$, $K_{0(V)} < K_0 < K_{0(R)}$, and $K_{0(S)} < K_0 < K_{0(T)}$, respectively. The (normalized) values of the Mohr-Coulomb parameters used in the analysis are: shear modulus $\frac{G}{\sigma_v} = 100$; Poisson's ratio $\nu = 0.3$; cohesion $\frac{c}{\sigma_v} = 0.5$; friction angle $\phi = 15°$, 30°, and 45°; dilation angle $\psi = 10°$; and the initial pore pressure $\frac{u_0}{\sigma_v} = 0.25$. These parameters together with the four reference $K_0$ values of $K_{0(V)}$, $K_{0(R)}$, $K_{0(S)}$, and $K_{0(T)}$ (related to $\frac{c}{\sigma_v}$ and $\phi$ via Eqs. (4a)-(4d)) are given in Table 1.

Figs. 4(a) and 4(b) show the normalized cylindrical cavity pressure-expansion curves (i.e., $\frac{\sigma_a}{\sigma_v}$ versus $\frac{a}{a_0}$) for the three different magnitudes of friction angle $\phi = 15°$, 30°, and 45° considered, with $K_0 = 0.8$ and 1.5, respectively. Note that according to Table 1, these two values of $K_0$ both fall into the first category of $K_{0(R)} < K_0 < K_{0(S)}$. However, for $\phi = 30°$ and $45°$, the two corresponding curves pertain to the solution case of $2\nu \geq 1 - \sin\phi$ (note: the critical condition of $2\nu = 1 - \sin\phi$ for a constant value of $\nu = 0.3$ is fulfilled at $\phi_{cr} = 23.58° < 30°$) with the stress path ending at point $\mathbf{X_u}$ in the sextant $\frac{3\pi}{2} < \theta < \frac{11\pi}{6}$; while for $\phi = 15°$, the cavity expansion curve corresponds to the case of $2\nu < 1 - \sin\phi$, accompanied by the occurrence of the intersection point $\mathbf{X_{pr}}$ (as marked in the two figures) before the ultimate stress state $\mathbf{X_u}$ located on the $s_r$ axis is reached. From Fig. 4, it can be clearly seen that the friction angle has a significant effect on the cavity responses; an increased value of $\phi$ results in stiffer cavity expansion curves and hence higher limit cavity pressure $\sigma_u$. Nevertheless, the calculated cavity expansion curve seems to have been only slightly influenced by $K_0$ when it varies from 0.8 to 1.5.

Fig. 5 presents the corresponding results for a relatively small value of $K_0 = 0.2$ that belongs to the second category of $K_{0(V)} < K_0 < K_{0(R)}$ (see Table 1), again for all the three values of friction angle involved. Remember that in this case scenario, the deviatoric stress path developed generally span two adjacent sextants of $\frac{7\pi}{6} < \theta < \frac{3\pi}{2}$ and $\frac{3\pi}{2} < \theta < \frac{11\pi}{6}$ in response to the continued expansion of the cavity. Therefore, the intersection point $\mathbf{X_{p\theta}}$ with the negative $s_\theta$ axis indeed occurs for each of the $\phi$ values, as indicated in the individual curves of the figure. It is noted that the friction angle however has a negligible influence on the calculated expansion radius $\frac{a_{p\theta}}{a_0}$ (where $a_{p\theta}$ denotes the cavity radius pertaining to the stress state $\mathbf{X_{p\theta}}$); the magnitudes of $\frac{a_{p\theta}}{a_0}$ are found to be 1.00281, 1.00280, and 1.00272 corresponding to $\phi = 15°, 30°,$ and $45°$. Also marked in Fig. 5 is the expanded cavity radius $\frac{a_{pr}}{a_0}$ corresponding to the intersection stress point $\mathbf{X_{pr}}$, which, as expected, occurs only for the case of $\phi = 15°$ with the satisfaction of $2\nu < 1 - \sin\phi$.

Fig. 6 further shows the calculated cavity expansion curves for the cases of $K_0 = 3.3$ (with $\phi = 15°$ and $30°$) and of $K_0 = 6$ (with $\phi = 45°$), to cover the last solution scenario of $K_{0(S)} < K_0 < K_{0(T)}$ for which the elastic-plastic transition point $\mathbf{X_{ep}}$ shifts downward to the segment $BT$. The reason that two different values of $K_0$ have been involved in this figure, rather than a single one as adopted in the previous two figures, is simply due to the fact that there exists no single value of $K_0$ satisfying $K_{0(S)} < K_0 < K_{0(T)}$ for all the $\phi$ cases considered, as is evident from Table 1. Once again, Fig. 6 clearly demonstrates that the cavity pressure $\frac{\sigma_a}{\sigma_v}$ increases profoundly with the increasing friction angle $\phi$. A further comparison of the results presented in this figure with those in Figs. 4 and 5 reveals that the calculated $\frac{\sigma_a}{\sigma_v}$ increases monotonously but moderately as $K_0$ increases from 0.2 to 6. Also marked in Fig. 6 is the $\frac{a_{pr}}{a_0}$ value corresponding to the intersection point $\mathbf{X_{pr}}$, which always occurs for the present category of $K_{0(S)} < K_0 < K_{0(T)}$.

Finally, the variations of the limit cavity pressure $\sigma_u$ with the friction angle $\phi$ (ranging from $10°$ to $50°$), for three different values of $K_0 = 0.2$, $0.8$, and $1.5$, are plotted in Fig. 7. It is evident that the effect of the friction angle is significant; an increase in $\phi$ from $10°$ to $50°$ will increase the ultimate cavity pressure $\sigma_u$ by almost two times. In contrast, the earth pressure coefficient $K_0$ appears to have only a minor influence on the calculated $\sigma_u$. Note that the solution procedure essentially switches from the case of $2\nu < 1 - \sin\phi$ to the case of $2\nu \geq 1 - \sin\phi$ across the critical value of $\phi_{cr} = 23.58°$ when the equal sign of the above inequality is fulfilled.

**Conclusions**

The graphical solution procedure proposed by Chen & Wang (2022), intended to rigorously solve the undrained cylindrical cavity expansion problem in non-associated Mohr-Coulomb soil with corner singularity, has been successfully extended in this paper to cover the general in situ stress conditions with arbitrary values of the earth pressure coefficient $K_0$. The development of the generalized solution involves three different categories of graphical formulation, depending on the relative position of the in situ stress state (or $K_0$ value) with respect to two reference points along the triaxial compression/extension lines in the deviatoric stress plane. Nevertheless, for each of the three graphical analysis categories, the deviatoric stress path turns out to be always comprised of a set of trackable, piecewise straight lines, as in the special hydrostatic in situ stress case with $K_0 = 1$. It is essentially such a desired feature of the stress path that makes possible the deduction of the cavity expansion responses in completely closed form, yet being free of the limitation of the intermediacy assumption for the vertical stress and of the mathematical difficulties involved in the treatment of corner singularity pertaining to the Mohr-Coulomb yield/potential surfaces. The removal of the former undesired assumption is of particular importance, since for soils with considerably high or low values of $K_0$, the vertical stress will remain the minor or major principal stress, immediately upon the commencement of yielding in the cylindrical cavity expansion.

The orientation of the straight-lined effective stress path, when located in the major sextant of $\frac{3\pi}{2} < \theta < \frac{11\pi}{6}$ with $\sigma'_r > \sigma'_z > \sigma'_\theta$, is found to be controlled merely by the two soil parameters of fiction angle and Poisson's ratio. It, however, depends on the third additional parameter of the dilation angle too if the stress path is situated in the sextants of $\frac{7\pi}{6} < \theta < \frac{3\pi}{2}$ or $-\frac{\pi}{6} < \theta < \frac{\pi}{6}$. For any arbitrary values of $K_0$ encountered, provided that the cavity is sufficiently expanded, the stress path will terminate exclusively either in the major sextant of $\frac{3\pi}{2} < \theta < \frac{11\pi}{6}$ ($2\nu \geq 1 - \sin\phi$) or on the projected $s_r$ axis ($2\nu < 1 - \sin\phi$). Parametric analyses show that the cavity expansion curve and the limit cavity pressure are significantly impacted by the soil friction angle, but only moderately by the earth pressure coefficient. The closed-form, graphical analysis-based solution derived in the current work can be regarded as a definitive and complete one for the undrained cavity expansion problem in the classical Mohr-Coulomb materials without the approximations and simplifications in previous solutions. This will be particularly valuable and beneficial in geotechnical practice as it pertains to the interpretations of pressuremeter tests in cohesive-frictional soils.

## Acknowledgements

The work reported in this paper was developed in a collaboration project between Aramco Services Company and Louisiana State University (Grant No. A-0123-2020) and partially funded by the Industrial Ties Research Subprogram of the Louisiana Board of Regents [Grant No. LEQSF(2019-22)-RD-B-01].

under drained condition. *Int. J. Numer. Anal. Mech. Geomech*. 42, No. 1, 132-142.

Chen, S. L. & Abousleiman, Y. N. (2022). A graphical analysis-based method for undrained cylindrical cavity expansion in modified cam clay soil. *Géotechnique*, https://doi.org/10.1680/jgeot.21.00172.

Chen, S. L. & Wang, X. (2022). A graphical method for undrained analysis of cavity expansion in Mohr-Coulomb soil. *Géotechnique*, https://doi.org/10.1680/jgeot.22.00088.

Chen, W. F. & Han, D. J. (1998). *Plasticity for structural engineers*. J. Ross Publishing. New York, NY, USA.

Collins, I. F., Pender, M. J. & Yan, W. (1992). Cavity expansion in sands under drained loading conditions. *Int. J. Numer. Anal. Mech. Geomech*. **16**, No. 1, 3-23.

Florence, A. L. & Schwer, L. E. (1978). Axisymmetric compression of a Mohr–Coulomb medium around a circular hole. *Int. J. Numer. Anal. Mech. Geomech*. **2**, No. 4, 367-379.

Gibson, R. E. & Anderson, W. F. (1961). In situ measurement of soil properties with the pressuremeter. *Civ. Engng Public Works Rev.* **56**, 615-618.

Hill, R. (1950). *The mathematical theory of plasticity*. Oxford, UK: Oxford University Press.

Hughes, J. M. O., Wroth, C. P. & Windle, D. (1977). Pressuremeter tests in sands. *Géotechnique* **27**, No. 4, 455-477.

Ladanyi, B. (1963). Expansion of a cavity in a saturated clay medium. *J. Soil Mech. Found. Div. ASCE* **89**, No. 4, 127-161.

Mair, R. J. & Muir Wood, D. M. (1987). *Pressuremeter testing: methods and interpretation*. UK: Butterworths.

Mántaras, F. M. & Schnaid, F. (2002). Cylindrical cavity expansion in dilatant cohesive-frictional materials. *Géotechnique* **52**, No. 5, 337-348.27

**Captions of tables and figures**

Table 1. $K_0$ values for reference points V, R, S, and T ($c/\sigma_v = 0.5$)

Fig. 1. Graphical representation of stress state/path in deviatoric plane for a soil element during cavity expansion process with $K_{0(R)} < K_0 < K_{0(S)}$

Fig. 2. Graphical representation of stress state/path in deviatoric plane for a soil element during cavity expansion process with $K_{0(V)} < K_0 < K_{0(R)}$

Fig. 3. Graphical representation of stress state/paths in deviatoric plane for a soil element during cavity expansion process with $K_{0(S)} < K_0 < K_{0(T)}$

Fig. 4. Cavity pressure-expansion curves for various friction angles under graphical solution category of $K_{0(R)} < K_0 < K_{0(S)}$: (a) $K_0 = 0.8$; (b) $K_0 = 1.5$

Fig. 5. Cavity pressure-expansion curves for various friction angles under graphical solution category of $K_{0(V)} < K_0 < K_{0(R)}$, $K_0 = 0.2$

Fig. 6. Cavity pressure-expansion curves for various friction angles under graphical solution category of $K_{0(S)} < K_0 < K_{0(T)}$, $K_0 = 3.3$ and $6$

Fig. 7. Limit cavity pressure in variation with friction angle for various $K_0$ values

**Table 1. $K_0$ values for reference points V, R, S, and T ($c/\sigma_v = 0.5$)**

| $\phi$ | $K_{0(V)}$ | $K_{0(R)}$ | $K_{0(S)}$ | $K_{0(T)}$ |
|---|---|---|---|---|
| 15° | −0.43 | 0.28 | 2.22 | 3.44 |
| 30° | −0.43 | 0.28 | 3.15 | 5.31 |
| 45° | −0.38 | 0.31 | 5.02 | 9.05 |

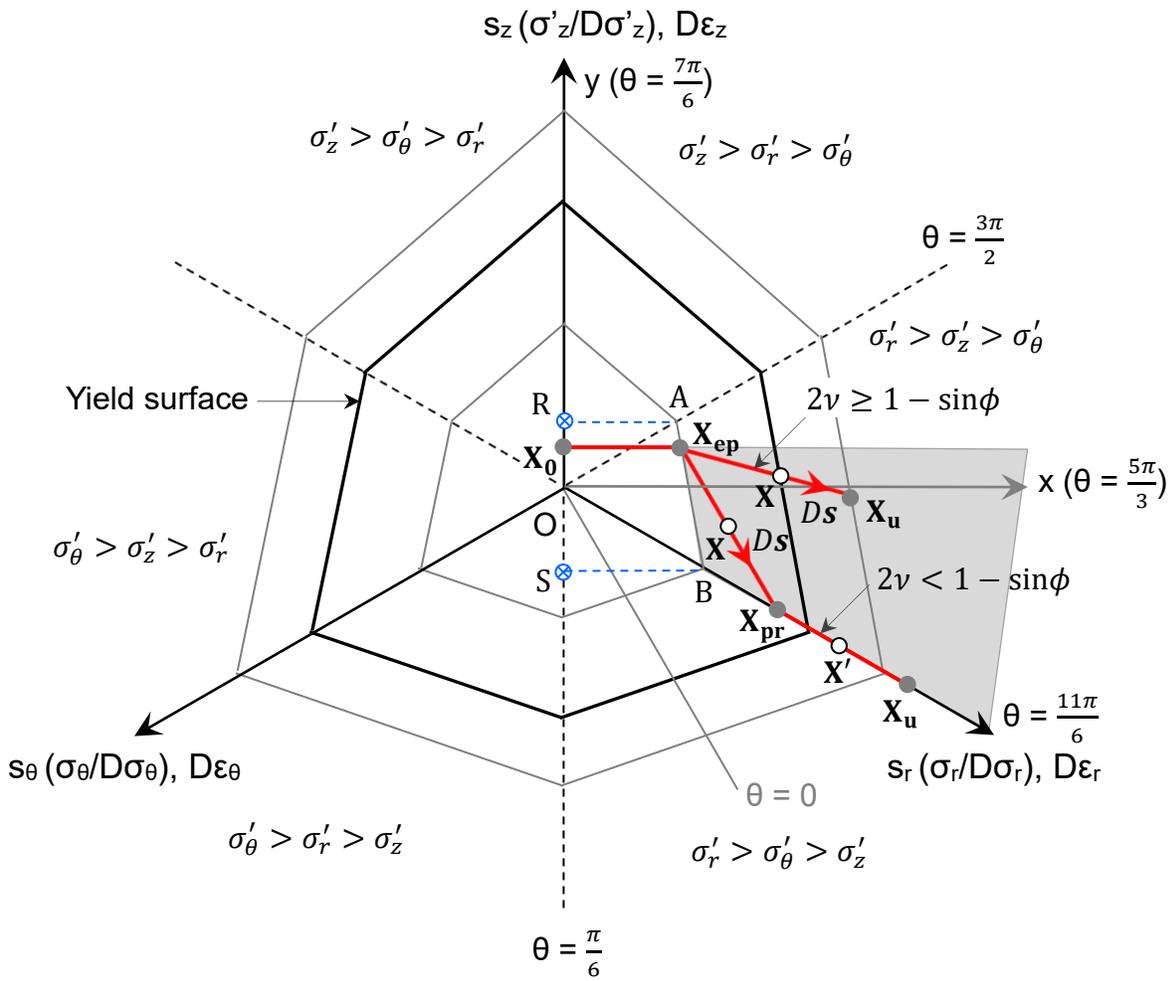

**Fig. 1.** Graphical representation of stress state/path in deviatoric plane for a soil element during cavity expansion process with $K_{0(R)} < K_0 < K_{0(S)}$

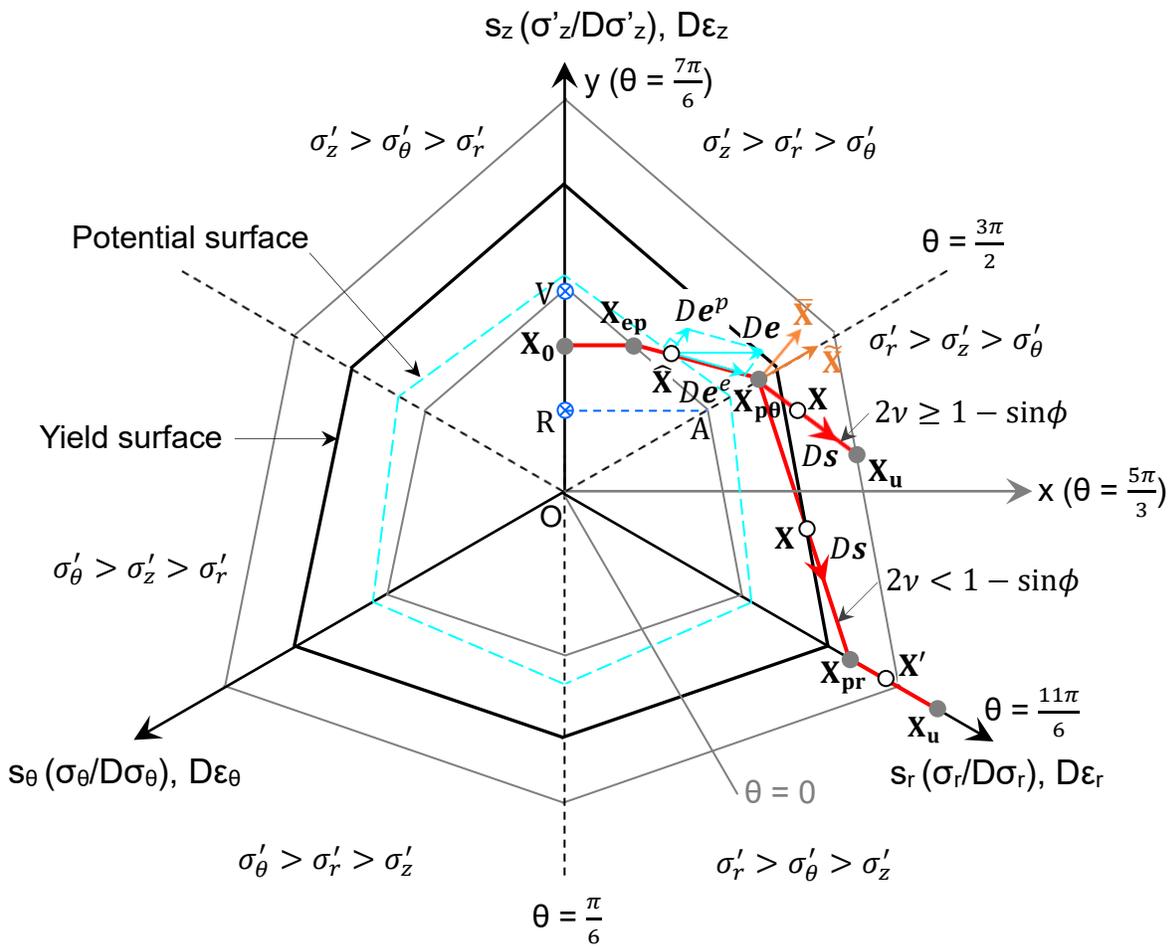

**Fig. 2. Graphical representation of stress state/path in deviatoric plane for a soil element during cavity expansion process with $K_{0(V)} < K_0 < K_{0(R)}$**

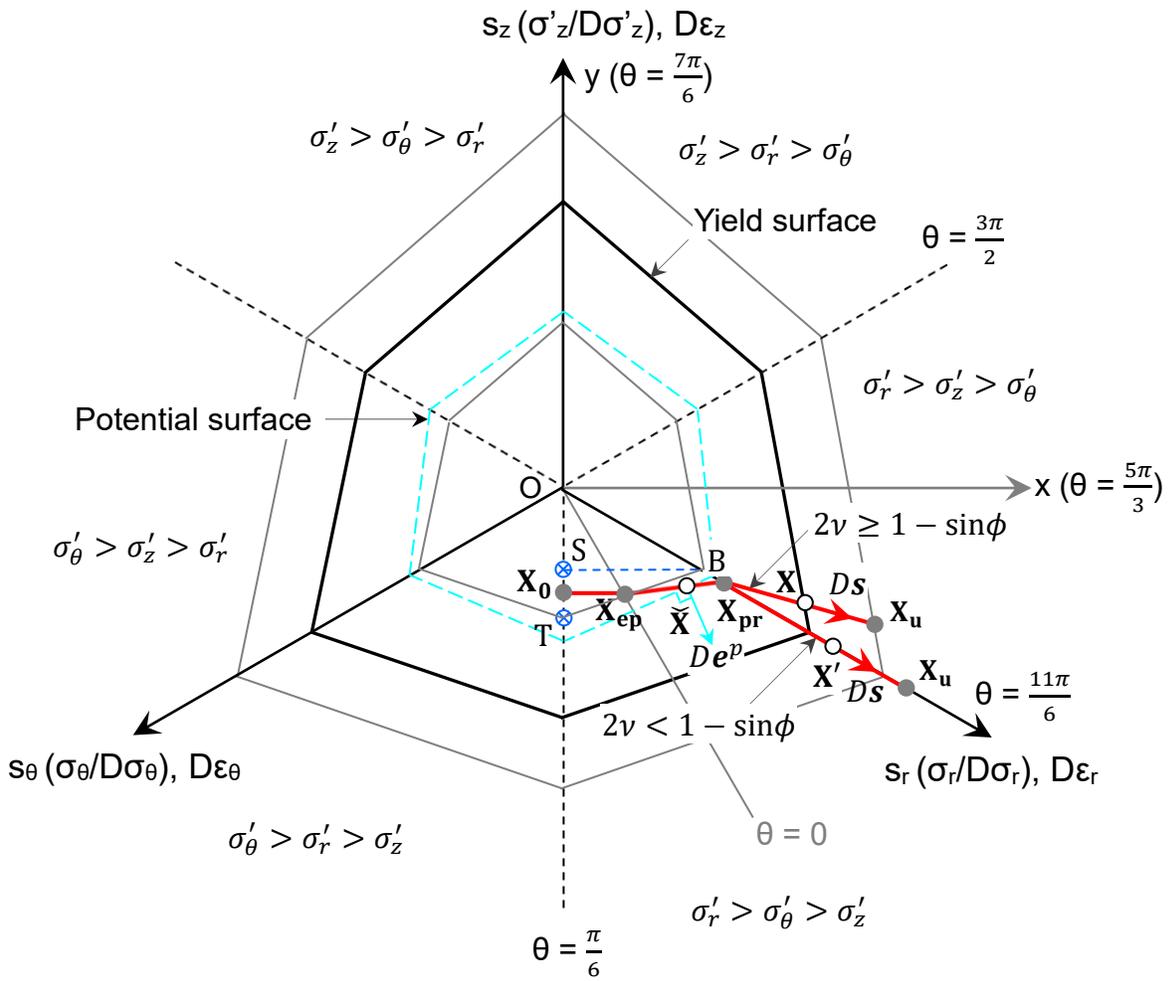

**Fig. 3.** Graphical representation of stress state/paths in deviatoric plane for a soil element during cavity expansion process with $K_{0(S)} < K_0 < K_{0(T)}$

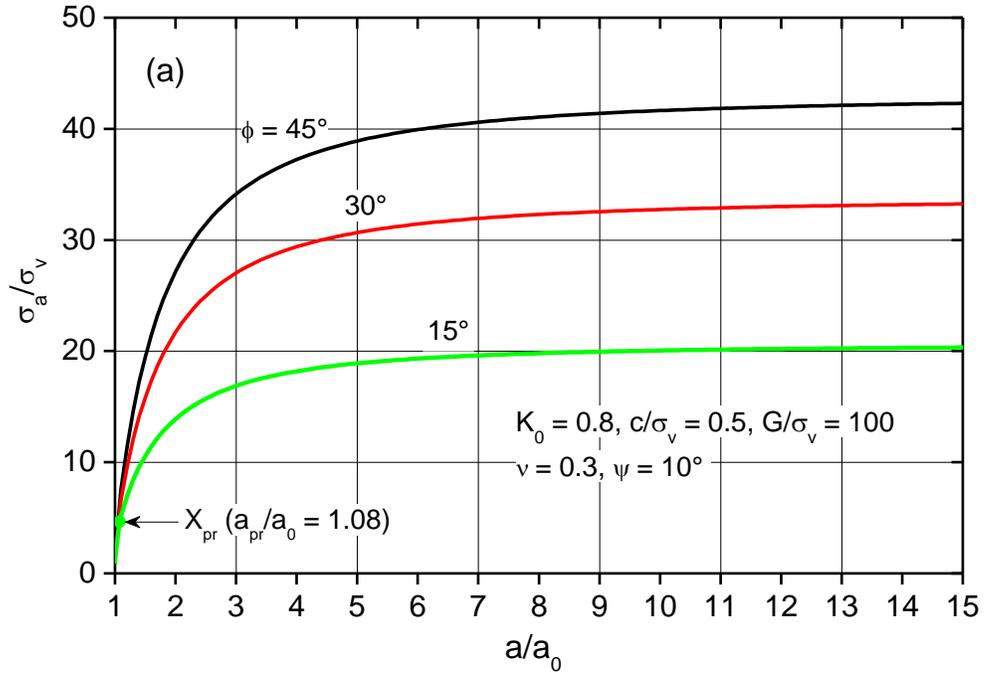

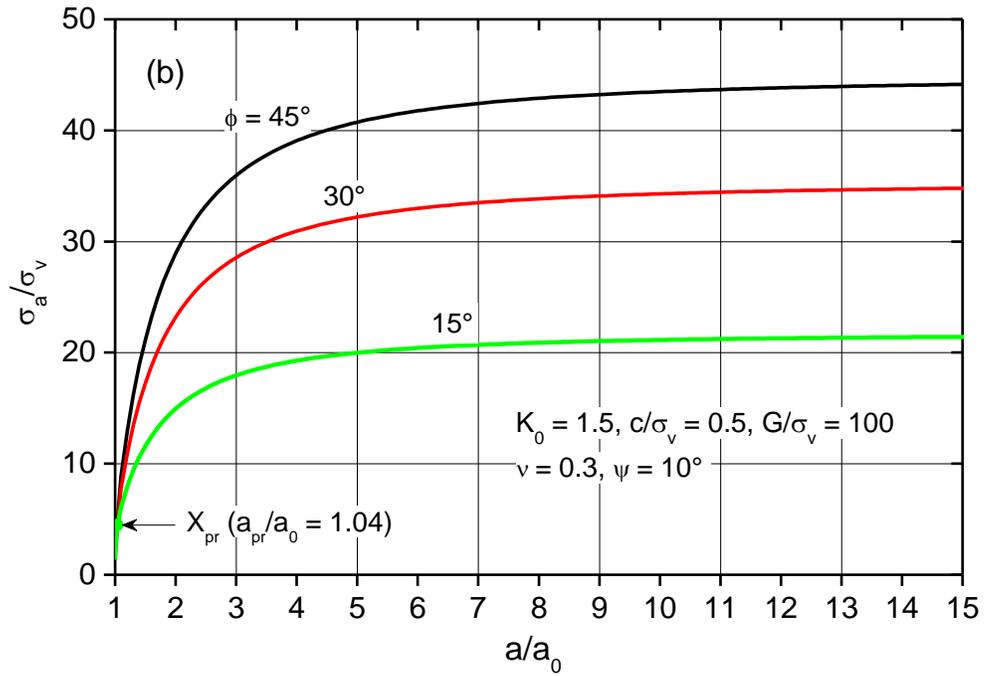

**Fig. 4.** Cavity pressure-expansion curves for various friction angles under graphical solution category of $K_{0(R)} < K_0 < K_{0(S)}$: (a) $K_0 = 0.8$; (b) $K_0 = 1.5$

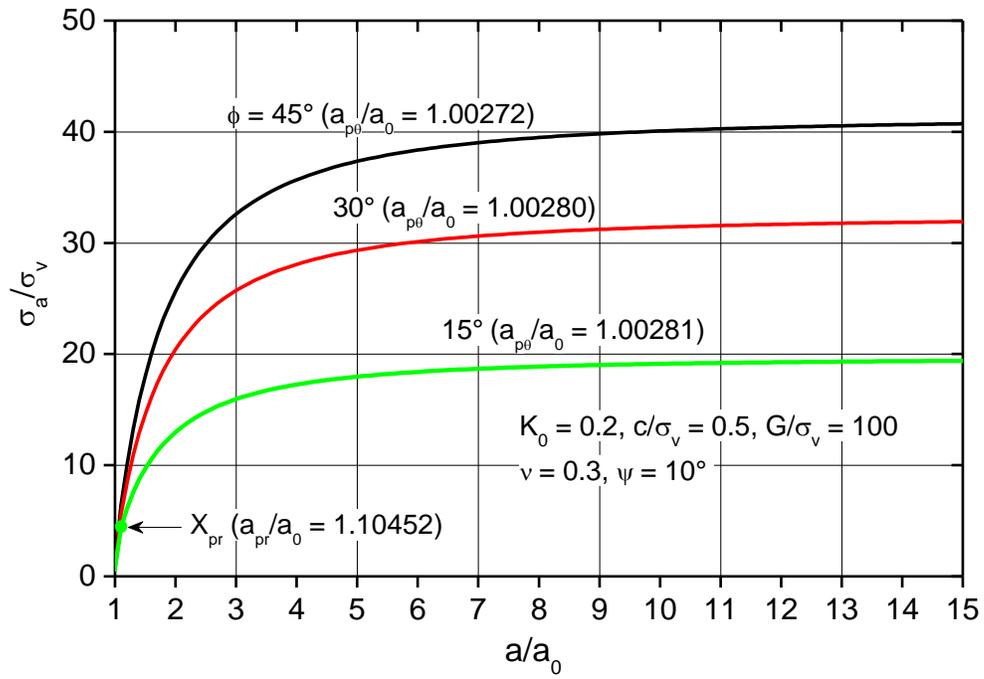

**Fig. 5.** Cavity pressure-expansion curves for various friction angles under graphical solution category of $K_{0(V)} < K_0 < K_{0(R)}$, $K_0 = 0.2$

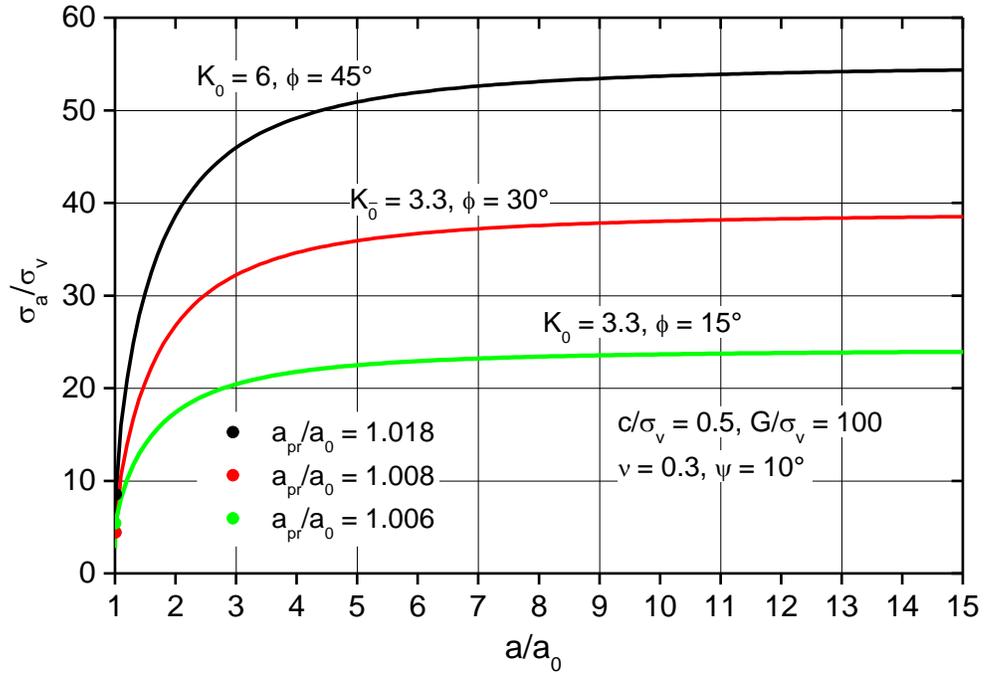

**Fig. 6. Cavity pressure-expansion curves for various friction angles under graphical solution category of $K_{0(S)} < K_0 < K_{0(T)}$, $K_0 = 3.3$ and 6**

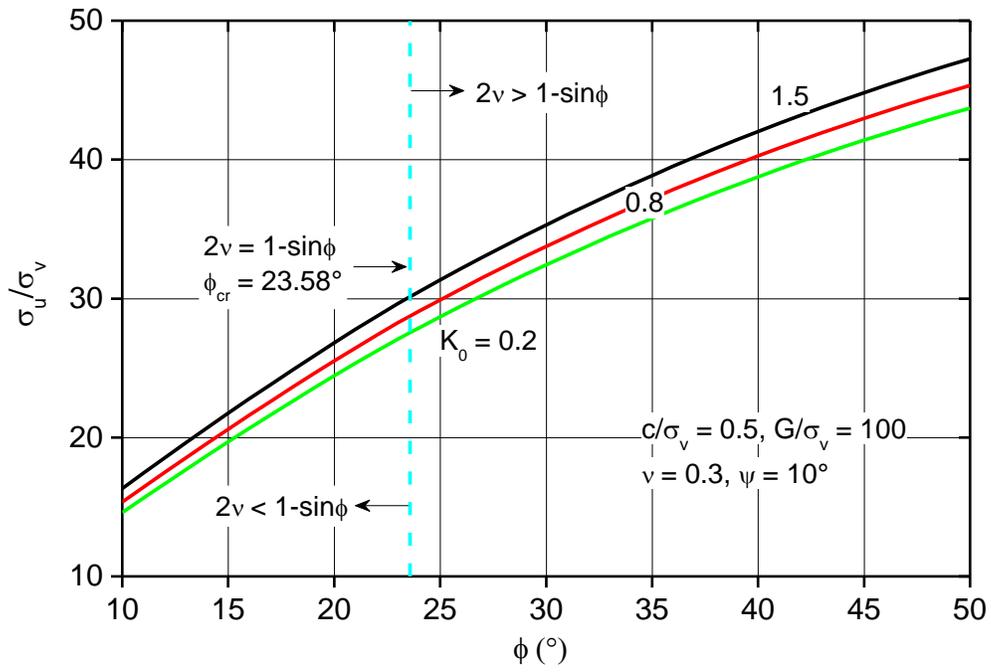

**Fig. 7. Limit cavity pressure in variation with friction angle for various $K_0$ values**